\documentclass[aps,twocolumn,showpacs]{revtex4}
\usepackage{CJK}
\usepackage{epsfig}
\def\bbox#1{\hbox{\boldmath${#1}$}}

\begin{document}
\begin{CJK*}{GBK}{song}

\title {$B_s$ Semileptonic Decays to $D_s$ and $D_s^*$ in
Bethe-Salpeter Method\footnote{Supported by NSFC under Contract No.10875032
and in part by SRF for ROCS, SEM.}}

\author{ZHANG Jin-Mei$^1$ and WANG Guo-Li$^1$}

\affiliation{$^1$Department of Physics, Harbin Institute of
Technology, Harbin 150001, China.}

\begin{abstract}

Using the relativistic Bethe-Salpeter method, the electron energy
spectrum and the semileptonic decay widths of $B^0_s\rightarrow
D^-_s \ell^+{\nu_\ell}$ and $B^0_s\rightarrow
D_s^{*-}\ell^+{\nu_\ell}$ are calculated. We obtained large
branching ratios, $Br(B_s\rightarrow D_se\nu_e)=(2.85\pm0.35)\% $
and $Br (B_s\rightarrow D_s^*e\nu_e)=(7.09\pm0.88)\%$, which can
be easily detected in the future experiment.

\end{abstract}

\pacs{13.20.He, 13.25.Ft, 13.25.Hw, 14.40.Lb, 14.40.Nd}

\maketitle

The light $B^0$ and $B^+$ meson decays have been precisely measured
in experiment and well studied in theories\cite{new}. Unlike $B^0$
and $B^+$ mesons, the heavier $B_s$ meson cannot be studied at the
$B$-factories operating at the $\Upsilon (4S)$ resonance. It can be
produced at the $\Upsilon (5S)$ resonance, Belle collaboration has
collected large number of $B_s$ samples in the year 2006 and 2008,
and they have presented a preliminary result of the inclusive
semileptonic branching fraction \cite{belle}: $Br(B^0_s\rightarrow
X^+ e^- \nu)=(10.9\pm1.0\pm0.9)\%$, and in the forthcoming LHCb
experiment, besides $B^0$ and $B^+$ mesons, an abundant number of
$B_s$ meson will be produced, so we have been provided a chance to
study the properties of $B_s$ meson and its various decay channels.

The physics of $B_s$ has become hot topic in recent years, for
example, the non-leptonic $B_s$ decays have been extensively studied
in the literature \cite{93,98,02,06,07,08}. Whereas the
investigation of semileptonic decays of $B_s$ to heavy meson is
relatively modest, for example, though there are the measurement of
$B_s\rightarrow D_s X \ell^+{\nu_\ell}$ \cite{PDG}, the pure decay
modes of $B_s\rightarrow D_s \ell^+{\nu_\ell}$ and $B_s\rightarrow
D_s^*\ell^+{\nu_\ell}$ have not been detected separately in
experiment. The semileptonic weak decays of heavy flavored mesons
are certainly interesting, it not only yield some of the most useful
information on the elements of the CKM matrix, $CP$-violation and
flavor-violation, but also increase our opportunities of discovering
new physics.

In this letter, we will study the exclusive semileptonic decays of
$B_s$ to $D_s$ and $D_s^*$ mesons based on the instantaneous
relativistic Bethe-Salpeter (BS) equation (Salpeter equation). Since
the BS equation is a full relativistic equation to describe bound
state, so we choose this method to consider the relativistic
corrections in $B_s$ decays. Another important thing in this decay
is that the $B_s$ is much heavier than the final state, there are
large recoil effect, this should be dealt with correctly. In our
calculation, besides the relativistic dynamic BS equation, we use
Mandelstam formalism \cite{S} to calculate the transition amplitude
with invariant variables as input, so we can correctly consider the
recoil effect, and give reliable calculations.

\begin{center}
\vspace{8mm}
\includegraphics[width=6.5cm]{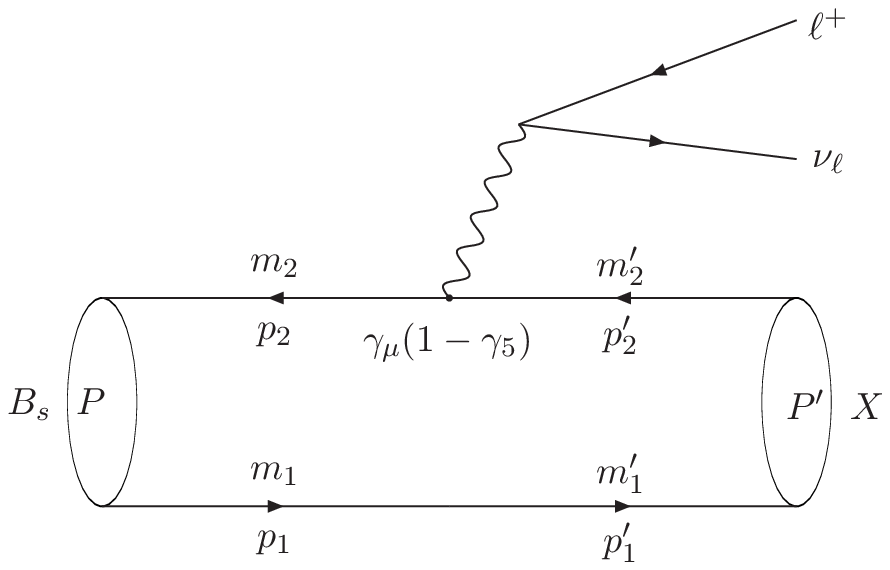}
\footnotesize
\begin{tabular}{p{6.5cm}}
\bf Fig.\,1. \rm Feynman diagram corresponding to the semileptonic
decays $B_s\rightarrow X+\ell^++{\nu_\ell}$
\end{tabular}
\end{center}

For the semileptonic decays $B_s\rightarrow X+\ell^++{\nu_\ell}$
(here $X$ denote as $D_s$ and $D_s^*$) shown in Fig.1, the
$T$-matrix element is:
\begin{eqnarray}
T=\frac{G_F}{\sqrt{2}}V_{ij}\bar{u}_{\nu_\ell}\gamma^{\mu}(1-\gamma_5)
v_{\ell}\langle X(P^\prime,\epsilon)|J_{\mu}|B_s(P)\rangle
\end{eqnarray}
where $V_{ij}$ is the CKM matrix element, $J_{\mu}$ is the charged
current responsible for the decays, $P$, $P'$ are the momenta of the
initial state $B_s$ and the final state $X$, respectively.

To evaluate the exclusive semileptonic differential decay rates of
$B_s$ meson, one needs to calculate the hadron matrix element of the
weak current $J_{\mu}$ sandwiched by the $B_s$ meson state as the
initial state and a single-hadron state of the concerned final
state, i.e., $\langle X(P^\prime,\epsilon)|J_{\mu}|B_s(P)\rangle$.
It is well known that the Mandelstam formalism \cite{S} is one of
proper approaches to compute the hadron matrix elements sandwiched
by the BS wave functions of the two bound-state, no matter how great
the recoil momentum carried by the elements will be. With the help
of this method and the instantaneous approximation \cite{Chang1},
the hadron matrix elements in the center of mass system of initial
meson can be written as \cite{C1,C2,C3}:
\begin{eqnarray}
\label{eq0001} &&\langle X(P^\prime)|J_\mu|B_s(P)\rangle\nonumber \\
&=&\int\frac{d{\bbox q}}{(2\pi)^3}Tr\left[
\bar{\varphi}^{++}_{_{P^\prime}}(\bbox
q^\prime)\frac{\not\!P}{M}{\varphi}^{++}_{_{P}}(\bbox
q)\gamma_{\mu}(1-\gamma_5)\right]
\end{eqnarray}
where $M$ is the mass of the initial meson $B_s$, $\bbox q$ and
$\bbox q^\prime$ are the relative three-momentum of the
quark-anti-quark in the meson $B_s$ and $X$, respectively, $\bbox
q^\prime=\bbox q-\frac{m_1^\prime}{m_1^\prime+m_2^\prime}\bbox
P^\prime$, $\bbox P^\prime$ is the three dimensional momentum of
finial hadron state $X$, ${\varphi}^{++}$ is the component of BS
wave function projected onto the positive energy for the relevant
mesons, and
$\bar{\varphi}^{++}_{_{P^\prime}}=\gamma_0({\varphi}^{++}_{_{P^\prime}})^+\gamma_0$.

Considering the semileptonic decay $B_s\rightarrow
D_s^*+\ell^++{\nu_\ell}$ as an example, for the initial state
pseudoscalar meson ($J^P=0^-$) $B_s$, the positive energy wave
function takes the general form \cite{d3}:
\begin{eqnarray}\label{a06}
\varphi_{0^-}^{++}(\bbox q)&=&\frac{M}{2}\left\{\left[f_1(\bbox
q)+f_2(\bbox q)
\frac{m_1+m_2}{\omega_1+\omega_2}\right]\right.\times\nonumber\\
&&\left[\frac{\omega_1+\omega_2}{m_1+m_2}+\frac{\not\!{P}}{M}
-\frac{\not\!{q_{_\bot}}(m_1-m_2)}{m_2\omega_1+m_1\omega_2}\right]+\nonumber\\
&&\left.\frac{\not\!{q_{_\bot}}\not\!P(\omega_1+\omega_2)}{M(m_2\omega_1+m_1\omega_2)}\right\}\gamma_5
\end{eqnarray}
where $q_{_\bot}=(0,\bbox q)$ and $\omega_i=\sqrt{m_i^2+\bbox q^2}$,
$f_i(\bbox q)$ are eigenvalue wave functions which can be obtained
by solving the full Salpeter equations. As for the finial state
vector meson ($J^P=1^-$) $D_s^*$, the positive energy wave function
takes the general form \cite{W}:
\begin{eqnarray}
\varphi_{1^-}^{++}(\bbox q^\prime)&=&\frac{1}{2}
\left[A\not\!\epsilon_{_\bot}^{\prime\lambda}+B\not\!\epsilon_{_\bot}^{\prime\lambda}\not\!{P^\prime}
+C(\not\!{q_{_\bot}^\prime}{\not\!\epsilon}_{_\bot}^{\prime\lambda}-q_{_\bot}^\prime\cdot\epsilon_{_\bot}^{\prime\lambda})\right.+\nonumber\\
&&D(\not\!{P^\prime}\not\!\epsilon_{_\bot}^{\prime\lambda}\not\!{q_{_\bot}^\prime}
-\not\!{P^\prime}q_{_\bot}^\prime\cdot\epsilon_{_\bot}^{\prime\lambda})+\nonumber\\
&&\left.q_{_\bot}^\prime\cdot\epsilon_{_\bot}^{\prime\lambda}(E+F\not\!{P^\prime}
+G\not\!{q_{_\bot}^\prime}+H\not\!{P^\prime}\not\!{q_{_\bot}^\prime})\right]
\end{eqnarray}
where $\epsilon$ is the polarization vector of meson, and $A,\ B,\
C,\ D,\ E,\ F,\ G,\ H$ are defined as:
\begin{eqnarray}
A&=&M^\prime\left[f_5(\bbox q^\prime)-f_6(\bbox q^\prime)
\frac{\omega_1^\prime+\omega_2^\prime}{m_1^\prime+m_2^\prime}\right]\nonumber\\
B&=&\left[f_6(\bbox q^\prime)-f_5(\bbox q^\prime)
\frac{m_1^\prime+m_2^\prime}{\omega_1^\prime+\omega_2^\prime}\right]\nonumber\\
C&=&\frac{M^\prime(\omega_2^\prime-\omega_1^\prime)}{m_2^\prime\omega_1^\prime+m_1^\prime\omega_2^\prime}\left[f_5(\bbox
q^\prime) -f_6(\bbox
q^\prime)\frac{\omega_1^\prime+\omega_2^\prime}{m_1^\prime+m_2^\prime}\right]\nonumber\\
D&=&\frac{\omega_1^\prime+\omega_2^\prime}{\omega_1^\prime\omega_2^\prime+m_1^\prime
m_2^\prime+{\bbox
q^\prime}^2}\times\nonumber\\
&&\left[f_5(\bbox q^\prime)-f_6(\bbox
q^\prime)\frac{\omega_1^\prime+\omega_2^\prime}{m_1^\prime+m_2^\prime}\right]\nonumber\\
E&=&\frac{m_1^\prime+m_2^\prime}{M^\prime(\omega_1^\prime\omega_2^\prime+m_1^\prime
m_2^\prime-{\bbox
q^\prime}^2)}\times\nonumber\\
&&\left\{M^{\prime2}\left[f_5(\bbox q^\prime)-f_6(\bbox q^\prime)
\frac{m_1^\prime+m_2^\prime}{\omega_1^\prime+\omega_2^\prime}\right]\right.-\nonumber\\
&&\left.{\bbox q^\prime}^2\left[f_3(\bbox q^\prime)+f_4(\bbox
q^\prime)
\frac{m_1^\prime+m_2^\prime}{\omega_1^\prime+\omega_2^\prime}\right]\right\}\nonumber\\
F&=&\frac{\omega_1^\prime-\omega_2^\prime}{M^{\prime2}(\omega_1^\prime\omega_2^\prime+m_1^\prime
m_2^\prime-{\bbox
q^\prime}^2)}\times\nonumber\\
&&\left\{M^{\prime2}\left[f_5(\bbox q^\prime)-f_6(\bbox q^\prime)
\frac{m_1^\prime+m_2^\prime}{\omega_1^\prime+\omega_2^\prime}\right]\right.-\nonumber\\
&&\left.{\bbox q^\prime}^2\left[f_3(\bbox q^\prime)+f_4(\bbox
q^\prime)
\frac{m_1^\prime+m_2^\prime}{\omega_1^\prime+\omega_2^\prime}\right]\right\}\nonumber\\
G&=&\left\{\frac{1}{M^\prime}\left[f_3(\bbox q^\prime)+f_4(\bbox
q^\prime)
\frac{m_1^\prime+m_2^\prime}{\omega_1^\prime+\omega_2^\prime}\right]\right.-\nonumber\\
&&\left.\frac{2f_6(\bbox
q^\prime)M^\prime}{m_2^\prime\omega_1^\prime+m_1^\prime\omega_2^\prime}\right\}\nonumber\\
H&=&\frac{1}{M^{\prime2}}\left\{\left[f_3(\bbox q^\prime)
\frac{\omega_1^\prime+\omega_2^\prime}{m_1^\prime+m_2^\prime}+f_4(\bbox
q^\prime)\right]\right.-\nonumber\\
&&\left.2f_5(\bbox q^\prime)
\frac{M^{\prime2}(\omega_1^\prime+\omega_2^\prime)}{(m_1^\prime+m_2^\prime)(\omega_1^\prime\omega_2^\prime+m_1^\prime
m_2^\prime+{\bbox q^\prime}^2)}\right\}
\end{eqnarray}
where $M^\prime$ is the mass of the final meson $D_s^*$ and
$E^\prime=\sqrt{M^{\prime2}+\bbox P^{\prime2}}$.

Now, using Eq. (2) and the formula as follows \cite{PDG}:
\begin{equation}
\Gamma=\frac{1}{8M}\frac{1}{(2\pi)^3}\int|T|^2dE_{\ell}dE^\prime
\end{equation}
the concerned widths of the semileptonic decays are calculated out
finally.

\begin{center}
\vspace{8mm}
\includegraphics[width=7.8cm]{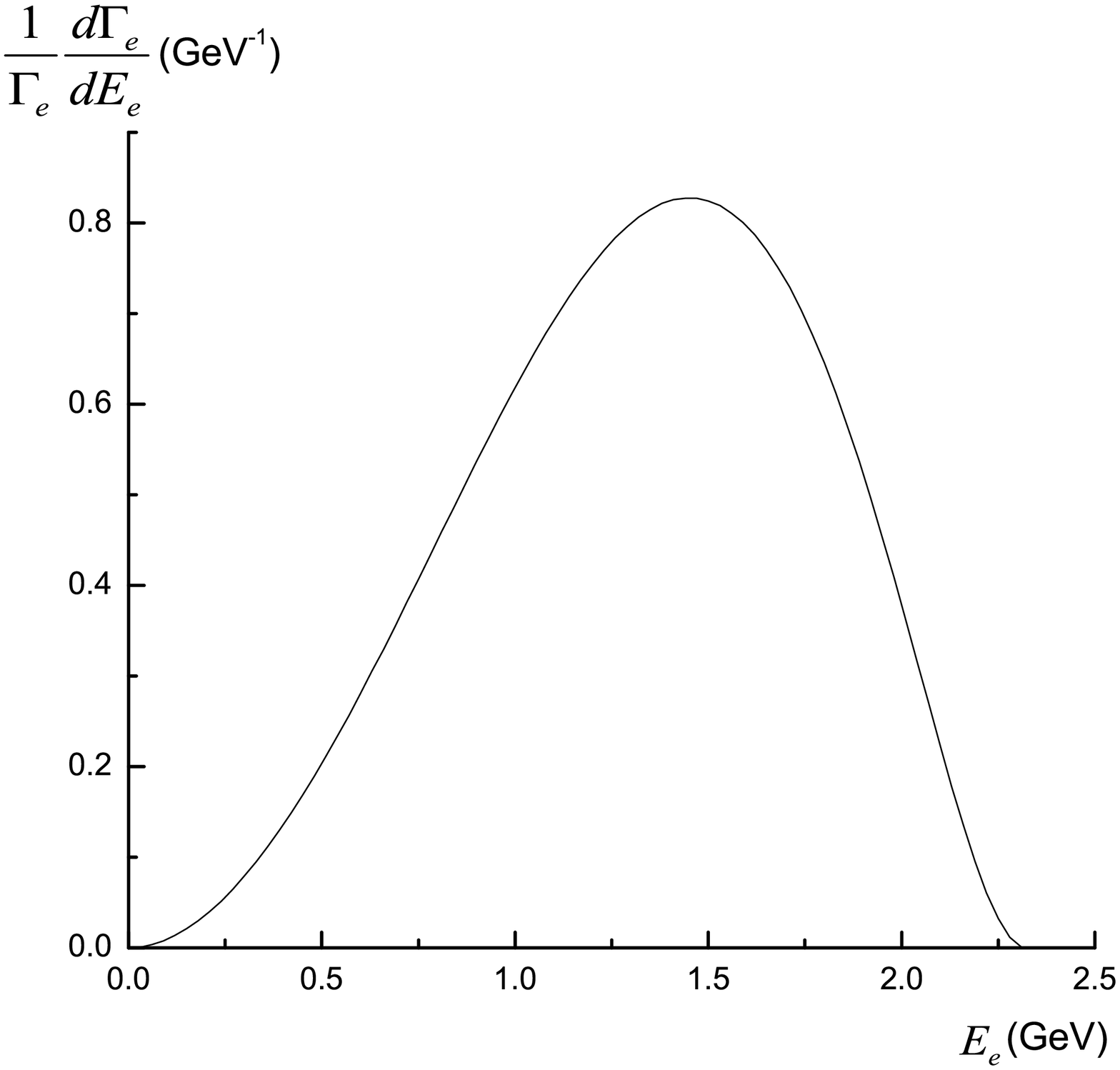}
\footnotesize
\begin{tabular}{p{7.8cm}}
\bf Fig.\,2. \rm The electron energy spectrum corresponding to the
semileptonic decays $B_s\rightarrow D_se^+{\nu_e}$
\end{tabular}
\end{center}
\begin{center}
\vspace{8mm}
\includegraphics[width=7.8cm]{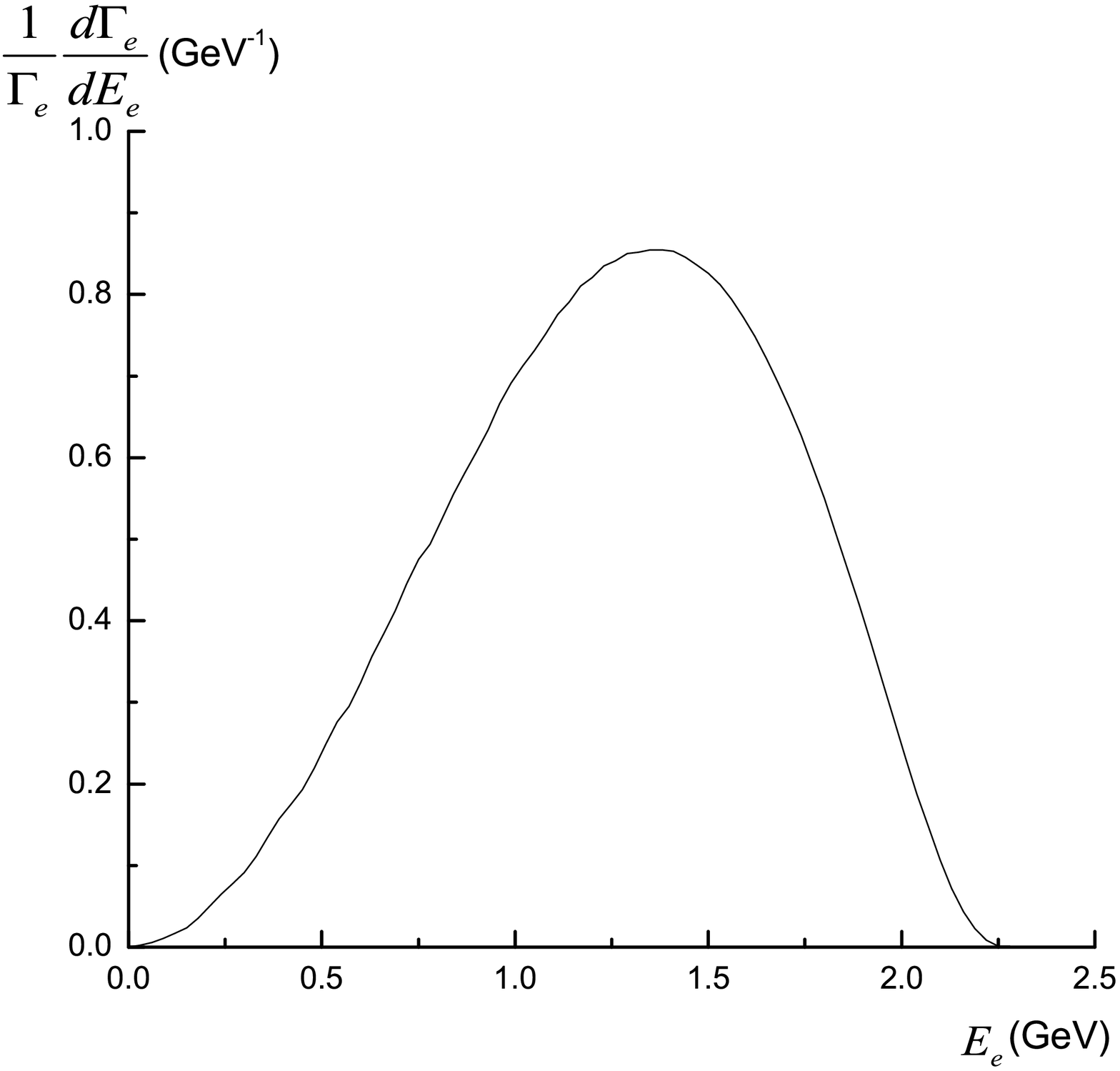}
\footnotesize
\begin{tabular}{p{7.8cm}}
\bf Fig.\,3. \rm The electron energy spectrum corresponding to the
semileptonic decays $B_s\rightarrow D_s^{*}e^+{\nu_e}$
\end{tabular}
\end{center}
In the numerical calculation, there are some parameters have to be
fixed. The input parameters for the masses of quarks are chosen as
follows: $a=e=2.7183,$ $\alpha=0.06$ GeV, $\lambda=0.1$ GeV$^2$,
$\Lambda_{QCD}=0.15$ GeV and $m_b=4.96$ GeV,\ $m_c=1.62$ GeV,\
$m_s=0.5$ GeV. For pseudoscalar mesons $B_s$, $D_s$ and vector meson
$D_s^*$, we choose $V_0=-0.194$ GeV, $V_0=-0.34$ GeV and $V_0=-0.15$
GeV. With the parameters, we obtained the masses: $M_{B_s}=5.3663$
GeV,\ $M_{D_s}=1.9685$ GeV and $M_{D_s^{*}}=2.1123$ GeV. The value
of the CKM matrix elements we used in this paper is: $V_{cb}=0.0412$
\cite{PDG}.

In Fig. 2 and Fig. 3 we show the spectra of the electron energy for
the decays $B_s\rightarrow D_se^+\nu_e$ and $B_s\rightarrow
D_s^*e^+\nu_e$, respectively.

Our numerical results of the exclusive semileptonic decay widths for
$D_s$ and $D_s^*$ final state are:
\begin{eqnarray}
\label{eq01} &&\Gamma(B_s\rightarrow D_se\nu_e)=(1.27\pm0.15)\times10^{-14}~\rm GeV\nonumber \\
&&\Gamma(B_s\rightarrow
D_s^*e\nu_e)=(3.17\pm0.39)\times10^{-14}~\rm GeV
\end{eqnarray}

Correspondingly, the branching ration of the two decays are:
\begin{eqnarray}
\label{eq02} &&Br(B_s\rightarrow D_se\nu_e)=(2.85\pm0.35)\%\nonumber \\
&&Br (B_s\rightarrow D_s^*e\nu_e)=(7.09\pm0.88)\%
\end{eqnarray}
so these two channels have large branching ratios, and with two
electronic particles in the final states, these two modes are easy
detected in experiment. Our results are close to the existing
theoretical results by Blasi $et~ al$ \cite{blasi}, which are
$\Gamma(B_s\rightarrow D_se\nu_e)=(1.35\pm 0.21)\times10^{-14}~\rm
GeV$ and $\Gamma(B_s\rightarrow D_s^*e\nu_e)=(2.5\pm
0.1)\times10^{-14}~\rm GeV $.

We also use this method to calculate the $B_s$ semileptonic decay
to $B$ and $B^*$, though these modes have a favor CKM matrix
element, $V_{us}=0.2255$ \cite{PDG}, they are deeply suppressed by
phase space. And the decay widths are:
\begin{eqnarray}
\label{eq02} &&\Gamma(B_s\rightarrow Be\nu_e)=0.85\times10^{-20} ~\rm GeV\nonumber \\
&&\Gamma(B_s\rightarrow B^*e\nu_e)=1.04\times10^{-21} ~\rm GeV
\end{eqnarray}
with the corresponding branching ratios:
\begin{eqnarray}
\label{eq03} &&Br (B_s\rightarrow
Be\nu_e)=0.19\times10^{-7}\nonumber \\
&&Br (B_s\rightarrow B^*e\nu_e)=0.23\times10^{-8}
\end{eqnarray}
which are too small to be measured.

In conclusion, we have calculated the decay widths of the
exclusive semileptonic $B_s$ decays to $D_s$ and $D_s^*$ mesons by
means of the instantaneous BS equation method. We find these two
modes have large branching ratios up to $10\%$ of the full decay
width, which consist with the preliminary result of Belle. But the
CKM matrix element favored channels, $B_s$ semileptonic decays to
$B$ and $B^*$, are suppressed deeply by the phase space and can
not be measured easily.

\end{CJK*}
\end{document}